# Anomalous stabilization of excitons by metallic proximity

Jeongkeun Song[1]*, Uksam Choi[2]*, Shan Lin[1]*, Du Li[3], Baekjune Kang[2], Li Yang[3], Ambrose Seo[4], Changhee Sohn[2]†, and Ho Nyung Lee[1]‡

[1] *Materials Science and Technology Division, Oak Ridge National Laboratory, Oak Ridge, TN 37831, USA.*
[2] *Department of Physics, Ulsan National Institute of Science and Technology, Ulsan, 44919, Republic of Korea.*

[3] *Department of Physics and Institute of Materials Science and Engineering, Washington University in St. Louis, St. Louis, MO, 63130, USA.*

[4] *Department of Physics and Astronomy, University of Kentucky, Lexington, KY 40506, USA.*

Correspondence to: chsohn@unist.ac.kr, hnlee@ornl.gov

**Abstract**

Metallic environments are generally expected to suppress excitons through strong dielectric screening, yet their influence can differ in composite systems where metallic and insulating regions coexist. Here, we investigate excitonic states in $Pd_xCu_{1-x}CrO_2$ thin films across a percolation-driven metal-insulator transition. Optical spectroscopy and many-body GW calculations show that $CuCrO_2$ hosts strongly bound excitons with a binding energy of about 489 meV. With increasing Pd substitution, the system approaches an insulator-to-metal transition near $x = 0.5$, consistent with the site-percolation threshold of a triangular lattice. In the pre-percolation regime, the excitonic resonance redshifts by 241 meV while the Tanguy continuum onset remains nearly unchanged, consistent with a substantial increase in exciton binding energy before metallization. An image-charge-based excitonic hydrogen model shows that isolated metallic regions can enhance electron-hole binding through image-charge interactions, whereas conventional screening is recovered once a continuous metallic network forms. Although this model provides a possible interpretation of the observed excitonic evolution, an alternative scenario in which metallic and excitonic responses originate from electronically distinct states and evolve independently cannot be excluded. These results reveal unusual metallic-excitonic coexistence near a percolation-driven metal-insulator transition and suggest nanoscale metallic proximity as a possible route for modifying excitonic interactions.

## Introduction

Excitons are central quasiparticles in semiconductors and insulators, governing optical absorption, emission and energy transport in materials relevant to optoelectronics and quantum technologies[1-4]. Their stability relies on sufficiently strong Coulomb attraction between electrons and holes. In most practical settings, however, nearby metallic environments weaken this attraction through dielectric screening and therefore reduce exciton binding energies[5-11]. This effect is well established in systems ranging from two-dimensional semiconductors to hybrid perovskites and molecular crystals[12,13]. As a result, metallic components, such as Au or Ag substrates, are usually regarded as detrimental to excitonic stability[11]. Consequently, substantial effort has focused on mitigating screening by spatially separating excitons from metallic regions or by engineering dielectric environments to preserve Coulomb interactions[3,14,15,16].

Despite this prevailing paradigm, the interaction between excitons and metallic environments at the nanoscale remains largely unexplored. When metallic regions are confined to nanometer length scales, the electrostatic response can deviate qualitatively from that of a bulk metal, potentially leading to nontrivial modifications of electron–hole interactions. Understanding this regime is particularly important for hybrid materials in which metallic and excitonic components coexist on comparable length scales, as well as for designing architectures that combine plasmonic and excitonic functionalities.

Here, we observe an unconventional evolution of excitonic states accompanying the development of metallic character in $Pd_xCu_{1-x}CrO_2$. Using epitaxial $Pd_xCu_{1-x}CrO_2$ thin films as a model system, we tune the system across a percolation-driven insulator-metal transition (IMT) by substituting Cu with Pd. The parent compound $CuCrO_2$, consisting of triangularly coordinated Cu layers separated by insulating $CrO_6$ octahedral layers, hosts strongly bound Wannier-Mott excitons with binding energies on the order of several hundred meV, arising from interlayer Cu–Cr electronic transitions.[17-20] (Figure 1a). As doping increases, isolated metallic regions emerge within the insulating matrix and eventually form a continuous metallic network near the percolation threshold. In the pre-percolation regime, we observe a pronounced redshift and sharpening of excitonic resonance. Within the Tanguy analysis, this spectral evolution corresponds to an increase in the exciton binding energy of up to 49%.

To explore whether nanoscale metallic regions could contribute to this anomalous behavior, we employ an electrostatic toy model based on the image-charge method. In this model, isolated metallic regions generate competing attractive and repulsive image-charge interactions. For certain nanoscale geometries, the attractive interaction can dominate, thereby strengthening electron-hole binding. These results suggest a possible framework for understanding exciton–metal interactions at the nanoscale and motivate the use of isolated metallic regions to modify excitonic properties.

**Results**

We first examined the optical response of $CuCrO_2$ thin films through spectroscopic ellipsometry. The imaginary part of the dielectric function ($\varepsilon_2$) exhibits a pronounced resonance near 3.4 eV, followed by a broad, step-like continuum at higher energies (Fig. 1b), characteristic of excitonic transitions[21-23]. The spectral feature around 3.4 eV can be decomposed into a primary excitonic resonance at 3.4 eV ($E_1$) and high-energy shoulder at 3.6 eV ($E_2$).

To quantify the excitonic properties, we model the $\varepsilon_2$ spectrum using the Tanguy model, which captures both the bound excitonic state and the broad continuum. Given that the exciton binding energy ($E_b$) is determined by the lowest-energy excitonic transition, we applied the Tanguy model exclusively to the $E_1$. Meanwhile, $E_2$ and charge-transfer transition (C) around 6 eV were modeled as the Lorentzian oscillators to capture overall spectral shape (see Supplementary Materials Note 1). The analysis yields an optical band gap ($E_g$) of 3.36 eV and the $E_b$ of approximately 489 meV for the $E_1$ resonance. This large $E_b$ significantly exceeds thermal energy at room temperature ($k_B T \approx 25$ meV), indicating robust excitonic stability.

To further understand the electronic origin of the exciton, we perform first-principles calculations. Figure 1c represents results from density functional theory (DFT) calculations that show the valence band is dominated by localized Cu 3*d* states, while the conduction band arises from more dispersive Cr 3*d* states (Fig. 1c)[19,20]. Notably, the band structure reveals an effective mass asymmetry: the conduction band exhibits a highly dispersive, parabolic profile (light electron mass), whereas the Cu 3*d*-derived valence band is remarkably flat. This lack of dispersion in the Cu 3*d* band dictates a heavy hole mass[24,25]. Many-body GW–BSE calculations reproduce the experimentally observed excitonic peak and reveal a redshift of ~500 meV relative to the non-interacting spectrum (Fig. 1d), consistent with the extracted $E_b$. The DFT calculation further exhibits two transitions at 4.0 eV and 4.26 eV, corresponding to the Cu 3*d*→Cr $t_{2g}$ and Cu 3*d*→Cr $e_g$ transitions. On the other hand, when electron-hole interactions are included, the calculated $\varepsilon_2$ reproduces the sharp excitonic resonance near 3.4 eV observed in Figure 1b. Comparing the calculated absorption spectra with and without electron-hole interactions indicates a redshift of 500 meV for the primary peak around 4 eV, which is consistent with the $E_b$ obtained from our Tanguy fit (Fig. 1d).

We next investigate how excitons evolve upon metallic regions in $Pd_xCu_{1-x}CrO_2$. As Pd content increases, the system undergoes a transition from an insulating to a metallic state[26]. Under the conventional macroscopic screening picture, the metallic state would be expected to weaken excitonic interactions as metallicity develops. X-ray diffraction (XRD) shown in Fig. s2 and energy dispersive spectroscopy (EDS) shown in Fig. s3, summarized in Supplementary Materials Note 2 and Note 3, confirm continuous structural evolution and compositional control. For $x \leq 0.3$, the system retains insulating behavior with well-defined excitonic features, while a broad low-energy resonance emerges below 1 eV, consistent with localized plasmonic response from metallic regions.

The dielectric response (Figs. 2a–d) reveals a clear percolation-driven transition. For $x \leq 0.3$, the system retains insulating behavior with well-defined excitonic features, while a broad low-energy resonance emerges below 1 eV, consistent with localized plasmonic response from metallic regions dispersed within the insulating $CuCrO_2$ matrix[27-29]. This macroscopic optical response can be simulated using effective medium approximation (EMA), which evaluates the effective dielectric function by averaging the individual constituents according to their respective volume fractions[29]. Notably, this low-energy hump is well-reproduced by the EMA (see Fig. s4 in Supplementary Materials Note 4). The EMA results, which model a macroscopic mixture of distinct metallic and insulating domains, confirm that the doped system functions as a heterogeneous composite rather than a homogeneous alloy[28,29].

At $x \approx 0.5$, the real part of the dielectric function becomes negative at low frequencies, and a Drude response develops in $\varepsilon_2$, indicating the formation of a continuous metallic network[27,30,31]. The onset of metallicity at $x \approx 0.5$ coincides with the theoretical site-percolation threshold for a two-dimensional triangular lattice[32,33] (Figure 2e), suggesting that Pd forms isolated metallic regions below this threshold and a connected network above it. In addition, electrical conductivity is observed for samples with $x \geq 0.5$, as shown in Fig. s5 in Supplementary Materials Note 5. This structural-optical correspondence suggests an evolution from isolated metallic regions to a continuous metallic network at $x \approx 0.5$, coincident with the disappearance of the excitonic resonance in $\varepsilon_2$ and the onset of strong metallic screening.

Despite the excitonic response remains well defined and is not quenched for $x < 0.5$, the excitonic response remains unchanged for $x < 0.5$. As Pd concentration increases from 0 to 0.4, the primary excitonic resonance peak $E_1$ systematically redshifts from 3.4 eV to approximately 3.2 eV (Fig. 3a). Analysis using the Tanguy model shows that the estimated optical band gap remains nearly unchanged, while the $E_b$ monotonically increases from 489 meV at $x = 0$ to about 730 meV at $x = 0.4$ (Fig. 3b). Estimated optical band gap is sum of $E_g$ and $E_b$ obtained from Tanguy model. The increase in the magnitude of $E_b$ (~241 meV) matches the observed redshift, indicating that the redshift arises from enhanced electron-hole attraction rather than estimated optical band gap renormalization. Note that the film thickness was determined by X-ray reflectivity (Fig. s6). The thickness information and optical model fitting conditions are provided in Table s3 of Supplementary Materials Note 6. We further tested possible uncertainties in the optical model fitting from the incidence angle and the ultrathin $CuCrO_2$ buffer layer by comparing dielectric functions extracted at 60° and 70°, using optical models with and without the 2 u.c. buffer layer, and testing fitting uniqueness with mean square error curves (Figs. s7–s9). These results show that the extracted dielectric functions remain unchanged to the incidence angle and buffer layer, supporting that the observed optical response has minimal influence on these fitting choices.

In addition to the $E_b$ shift, the excitonic resonance becomes sharper and more intense with increasing Pd content (Fig. 3c). To isolate the intrinsic excitonic contribution, we evaluated the full width at half-maximum (FWHM) of the Tanguy fit, alongside the volume normalized intensity (VNI) of the $\varepsilon_2$ maximum. The VNI effectively compensates for the reduced volume fraction of the insulating domain by normalizing the raw peak intensity by a factor of $1/(1-x)$. As the Pd concentration increases, the FWHM systematically decreases while the VNI increases, providing evidence of excitonic enhancement.

This simultaneous exciton peak redshift and sharpening rules out the possibility of disorder-induced broadening or defect-related exciton trapping. If the redshift were driven by structural disorder or defect states introduced by Pd doping, the optical spectra would be expected to exhibit inhomogeneous peak broadening and splitting due to the coexistence of free and trapped excitons[34-36]. Moreover, the second derivative of $\varepsilon_2$ shows that the $E_1$ peak shifts to lower energy, while the higher energy charge-transfer feature remains nearly unchanged in the pre-percolation regime (Fig. s10 in Supplementary Materials Note 9). This supports that the redshift of $E_1$ peak reflects stabilization of the $CuCrO_2$-derived exciton, rather than a reconstruction of the band structure. These observations indicate that the excitonic resonance undergoes an anomalous evolution in the pre-percolation regime in contrast to conventional expectations based on dielectric screening. Another possible explanation would be mass enhancement of electrons and holes by doping. However, it is highly unlikely that $PdCrO_2$ has much larger bandwidth than $CuCrO_2$. Specifically, $CuCrO_2$ is known to have heavy holes with an effective mass of $m_h^* = 4 - 6\ m_0$ [Ref. [37]] while $PdCrO_2$ exhibits a lighter hole mass of $m_h^* = 1 - 1.5\ m_0$ [Ref. [38,39]], where $m_0$ is the free-electron mass.

To elucidate the possible origin of the anomalous behavior of the $E_b$ enhancement, we employed an excitonic hydrogen model based on the image-charge method[40-42]. Details of the model calculation are described in the Supplementary Materials Note 10. As demonstrated by our previous optical-structural analysis, metallic regions are embedded within the insulating $CuCrO_2$ matrix. Because these embedded metallic regions have no direct contact with an external charge reservoir, there is no net charge influx. In this sense, they are modeled as isolated and neutral metal spheres of radius $R$. Furthermore, given the remarkably flat valence band confirmed by our DFT calculations, the hole is treated as a static positive charge. In this model, the induced charge distribution on the metal sphere is simplified as two discrete point charges: a near-surface image-charge and a neutralizing center image-charge[43] (Figure 4a). Consequently, three distinct electrostatic interactions emerge for a single exciton: self-interaction (between a charge and its own image), cross-interaction (between a charge and the counter-charge's image), and center-interaction (between charges and the center image charges). Incorporating all these interactions under the heavy-hole approximation, the perturbed Hamiltonian for the exciton is defined as $H = H_0 + U_{ind}$, where $H_0 = -\frac{\hbar^2}{2m_e}\nabla^2 - \frac{e^2}{4\pi\varepsilon_0\varepsilon_r r}$ is the unperturbed exciton Hamiltonian, and $U_{ind}$ represents the total image-charge-induced perturbation (the sum of self-, cross-, and center-interactions). The perturbed Hamiltonian was solved using the variational method, with a hydrogen like 1$s$ trial wave function.

Based on this electrostatic model, we reveal that the geometric scale of the metallic regions can have entirely opposite effects on excitons: stabilizing or quenching. In the macroscopic bulk limit ($R >> a_X$), the attractive contribution vanishes, and the repulsive interaction dominates, recovering conventional screening behavior. In contrast, when the metallic region size is comparable to the exciton ($R \sim a_X$), the attractive interaction can dominate, leading to an increase in exciton binding energy by up to 7% (Figs. 4b, c). Under these conditions, metallic regions effectively act as electrostatic confinement centers that enhance electron–hole attraction.

To capture the physical intuition for how image-charges affect the exciton, we assume the electron and hole are placed at an identical distance $d$ from the sphere center ($r_e \sim r_h = d$), where $r_e$ and $r_h$ are distances from center of the sphere to respective charges. The center-interaction, $U_{cent}$, arises from the neutralizing image-charges $q_i^{cent} = +q_i \frac{R}{r_i}$, ($i = e, h$) at the sphere center. The explicit form of $U_{cent}$ is formulated as:

$$U_{cent} = \frac{1}{4\pi\varepsilon}\left[\frac{1}{2}\frac{q_e q_e^{cent}}{|r_e|} + \frac{1}{2}\frac{q_e q_h^{cent}}{|r_e|} + \frac{1}{2}\frac{q_h q_e^{cent}}{|r_h|}\right] \approx \frac{e^2 R}{8\pi\varepsilon}\left[\frac{1}{d^2} - \frac{2}{d^2}\right] = -\frac{e^2 R}{8\pi\varepsilon d^2} \quad (1)$$

This negative sign proves that the net center image-charge interaction is attractive ($\propto \left[-\frac{1}{d^2}\right]$), actively pulling the electron cloud toward the heavy hole. Meanwhile, the near-surface image interaction (sum of self- and cross-interaction), $U_{int}$, originates from the image-charges $q_i' =$ $-q_i \frac{R}{r_i}$) located at a distance $r_i' = \frac{R^2}{r_i}$ from the center of metal sphere. This interaction can be formulated as:

$$U_{int} = \frac{1}{4\pi\varepsilon}\left[\frac{1}{2}\frac{q_e q_e'}{|r_e - r_e'|} + \frac{1}{2}\frac{q_e q_h'}{|r_e - r_h'|} + \frac{1}{2}\frac{q_h q_e'}{|r_h - r_e'|}\right] = \frac{e^2 R}{8\pi\varepsilon(d^2 - R^2)} \tag{2}$$

Thus, the near-surface image interaction is repulsive ($\propto +\frac{1}{(d^2-R^2)}$) acting to dissociate the exciton. As shown in Eqs. (1) and (2), two distinct interactions compete on comparable energy scales. At sufficiently small distance between the exciton and the metal sphere, the attractive $U_{cent}$ can dominate over the repulsive $U_{int}$. This dominance of $U_{cent}$ leads to the $E_b$ enhancement in the microscopic regime.

The macroscopic regime can be derived by taking large-radius limit ($R \to \infty$) of previous equations. By defining $z$ as the distance from the metal surface ($d = R + z$), the attractive center image-charge potential vanishes:

$$U_{cent} \propto -\frac{R}{(R+z)^2} \to 0 \tag{3}$$

In contrast, the repulsive near-surface image interaction converges exactly to the classical infinite-plane limit:

$$U_{int} \propto \frac{R}{(R+z)^2 - (R)^2} = \frac{R}{2Rz + z^2} \to \frac{1}{2z} \tag{4}$$

Consequently, in this macroscopic limit, the net electrostatic potential weakens the electron-hole attraction ($\Delta E_b < 0$), converging to conventional macroscopic metallic screening typically observed at metal semiconductor interfaces. Note that this image-charge calculation is based on a single metallic region. The 7% enhancement of $E_b$ is therefore not intended to reproduce the full experimental increase in $E_b$ , but to show quantitatively that one nanoscale metallic region can enhance electron-hole binding under appropriate geometric conditions.

This metal size-dependent electrostatic analysis provides a possible interpretation of the percolation-driven excitonic evolution in $Pd_xCu_{1-x}CrO_2$. Using $a_X^* = a_0\varepsilon_r/\mu$, where $a_X^*$ is the exciton size, $a_0$ is the Bohr radius, $\varepsilon_r$ is the effective relative dielectric constant[44], and $\mu$ is the effective mass of the hole[37] estimated from reported $CuCrO_2$ study, we obtain $a_X^* \approx 2$ Å. This estimate suggests that the exciton is sufficiently localized to potentially experience local electrostatic interactions associated with nanoscale metallic regions. Within our image-charge model, such interactions can enhance electron–hole binding when the characteristic metallic length scale is comparable to that of the exciton, whereas the conventional screening limit is recovered as the metallic region becomes extended.

We emphasize, however, that this model establishes the physical possibility of metallic-region-induced enhancement rather than uniquely identifying its microscopic origin in $Pd_xCu_{1-x}CrO_2$. In particular, the calculated enhancement for a single metallic region is smaller than the experimentally inferred change in binding energy, and the present measurements do not directly establish microscopic coupling between the electronic states responsible for the metallic response and those forming the Cu–Cr excitonic transition. An alternative scenario in which Pd-derived low-energy metallic states and Cu–Cr excitonic states are electronically distinct and evolve largely independently therefore cannot be excluded. When Pd concentration reaches the percolation threshold ($x > 0.5$), these isolated metallic regions merge into a continuous metallic network. In this regime, metallic screening quenches exciton resonance. This observation is consistent with the recovery of conventional metallic screening, although it does not by itself establish screening as the unique cause of exciton quenching.

In summary, we investigate the anomalous stabilization of excitons in $Pd_xCu_{1-x}CrO_2$ across a percolation-induced metal–insulator transition. While conventional metallic screening suppresses excitonic states, the presence of isolated metallic regions enhances electron–hole binding, leading to a substantial increase in exciton binding energy prior to metallization. This behavior is captured by an image-charge-based electrostatic model and reverses once continuous metallic network forms. These findings highlight an unusual coexistence of evolving metallic and excitonic responses near a percolation transition and suggest that nanoscale metallic environments may offer opportunities for modifying excitonic interactions in hybrid materials.

## Methods

### Film fabrication and structural characterization

The $Pd_xCu_{1-x}CrO_2$ thin films were epitaxial synthesized on (0001)-oriented sapphire substrates by pulsed laser deposition. Before the thin film growth, commercially available $Al_2O_3$ substrates (CrysTec, Germany) were annealed at 1100 °C for 1 h. To avoid $Cr_2O_3$ impurity, 2 u.c. thick $CuCrO_2$ was used as a buffer layer. The oxygen pressure and temperature were maintained at 100 mTorr and 600 °C, respectively. The repetition rate and energy fluence of the KrF excimer laser ($\lambda$ = 248 nm) were fixed at 5 Hz and 1.5 J/cm$^2$, respectively. To achieve desired Pd-doping, we used two targets ($CuCrO_2$ and $PdCrO_2$) during the growth process by varying the pulse ratio. XRD with a four-circle high-resolution X-ray diffractometer (X'Pert Pro, PANalytical; Cu $K\alpha_1$ radiation) was used to confirm crystal structure of films.

### Ellipsometry and optical properties of $Pd_xCu_{1-x}CrO_2$

Spectroscopic ellipsometry was performed using an M-2000 ellipsometer (J.A. Woollam Co.) to obtain the real ($\varepsilon_1$) and imaginary ($\varepsilon_2$) parts of the dielectric function. Prior to all measurements, the instrument was calibrated at 300 K using a 25 nm Si/$SiO_2$ wafer to minimize the mean square error (MSE) of the optical signal. Ellipsometric spectra were acquired at an incident angle of 70 °, yielding $\Psi$ and $\Delta$ over the photon-energy range of 0.74 - 6.46 eV.

To prevent back-reflection signals originating from the rear surface of the sample, the specimens were mounted on an oxygen-free copper cone during measurements. Infrared ellipsometry was performed using an IR-VASE MARK 2 (J.A. Woollam Co.) over the photon-energy range of 0.04 - 0.81 eV. For analysis of plasmon-resonance peaks in Pd-doped $CuCrO_2$, the spectral window 0.2 eV-0.74 eV was used. Measurements were conducted at an incident angle of 70°. To reduce noise levels, data were accumulated for a total acquisition time of 3 h.

**Extraction of optical properties and excitonic modelling of $CuCrO_2$**

The raw Ψ-Δ data were converted into $\varepsilon_1$ and $\varepsilon_2$ using the WVASE analysis software. Specifically, to extract dielectric functions of Pd-doped $CuCrO_2$ thin films, a c-$Al_2O_3$ (0001) single-crystal substrate was first measured independently. The dielectric functions of Pd-doped $CuCrO_2$ thin films were obtained using a bilayer model provided in WVASE program, consisting of a film layer atop a c-$Al_2O_3$ substrate.

Analysis of the excitonic optical response was conducted using the General Oscillator (GenOsc) within the WVASE software. The experimentally obtained dielectric functions were modeled using a combination of the Tanguy model and Lorentz oscillators. The fitting parameters in Tanguy model include the exciton oscillator strength, optical band gap, exciton binding energy, and the broadening parameter. The $E_1$ peak was fitted using the Tanguy model because $E_b$ was extracted with respect to the continuum onset of the lowest Cu-Cr interlayer excitation. Since the $E_2$ peak also arises from a Cu-Cr excitation, the $E_2$ peak should share the same continuum. However, assigning an independent Tanguy oscillator to $E_2$ would introduce a separate continuum. Therefore, only $E_1$ was treated with the Tanguy model, while $E_2$, higher-energy features, and Urbach tails were described using Lorentz oscillators.

**Theoretical calculation and modelling**

DFT calculations for $CuCrO_2$ were performed using the PBE exchange–correlation functional, as implemented in the Quantum ESPRESSO package[45]. A wavefunction cutoff of 60 Ry and a charge density cutoff of 240 Ry were adopted. Hubbard parameters U = 5.0 eV and 4.0 eV[46] were used for the copper and chromium atoms, respectively. Our simulations indicate that the magnetic order does not significantly affect the electronic band gap and primary optical properties. Consequently, we focus on the ferromagnetic order in the following many-body perturbation theory calculations for $CuCrO_2$.

We calculated the quasiparticle (QP) energies within the GW approximation. The GW calculations were performed using the BerkeleyGW package[47]. All GW calculations reported in this work are perturbative, i.e., the single-shot $G_0W_0$. The Hybertsen−Louie generalized plasmon-pole model was used to treat the frequency dependence of the dielectric function[48]. The static remainder approximation[49] was used in the evaluation of self-energy for faster convergence. The Bethe-Salpeter equation (BSE) was employed to obtain excitonic effects and optical absorption spectra[50]. The coarse k grid was set to be $6 \times 6 \times 2$ for calculating the dielectric function and QP energies, and it was then interpolated to a fine k grid of $12 \times 12 \times 4$ for computing the electron-hole interaction kernel and solving the BSE. Gaussian curves with an energy resolution of 0.10 eV were adopted to broaden the exciton peaks in the calculated absorption spectrum.

## Reference

1 Sanvitto, D. & Kéna-Cohen, S. The road towards polaritonic devices. *Nature materials* **15**, 1061-1073 (2016).

2 Wang, G. *et al.* Colloquium: Excitons in atomically thin transition metal dichalcogenides. *Reviews of Modern Physics* **90**, 021001 (2018).

3 Mueller, T. & Malic, E. Exciton physics and device application of two-dimensional transition metal dichalcogenide semiconductors. *npj 2D Materials and Applications* **2**, 29 (2018).

4 Mak, K. F. & Shan, J. Photonics and optoelectronics of 2D semiconductor transition metal dichalcogenides. *Nature Photonics* **10**, 216-226 (2016).

5 Langreth, D. C. Approximate screening functions in metals. *Physical Review* **181**, 753 (1969).

6 Pirker, L., Honolka, J., Velický, M. & Frank, O. When 2D materials meet metals. *2D Materials* **11**, 022003 (2024).

7 Schöne, W.-D. & Ekardt, W. Transient excitonic states in noble metals and Al. *Physical Review B* **65**, 113112 (2002).

8 Wang, F. *et al.* Observation of excitons in one-dimensional metallic single-walled carbon nanotubes. *Physical review letters* **99**, 227401 (2007).

9 Spataru, C. D. Electronic and optical gap renormalization in carbon nanotubes near a metallic surface. *Physical Review B—Condensed Matter and Materials Physics* **88**, 125412 (2013).

10 Klots, A. R. *et al.* Controlled dynamic screening of excitonic complexes in 2D semiconductors. *Scientific reports* **8**, 768 (2018).

11 Park, S. *et al.* Direct determination of monolayer MoS2 and WSe2 exciton binding energies on insulating and metallic substrates. *2D Materials* **5**, 025003 (2018).

12 Filip, M. R., Qiu, D. Y., Del Ben, M. & Neaton, J. B. Screening of excitons by organic cations in quasi-two-dimensional organic–inorganic lead-halide perovskites. *Nano letters* **22**, 4870-4878 (2022).

13 Shunak, L., Adeniran, O., Voscoboynik, G., Liu, Z.-F. & Refaely-Abramson, S. Exciton modulation in perylene-based molecular crystals upon formation of a metal-organic interface from many-body perturbation theory. *Frontiers in chemistry* **9**, 743391 (2021).

14 Itzhak, R. *et al.* Exciton Manipulation via Dielectric Environment Engineering in 2D Semiconductors. *ACS Applied Optical Materials* **3**, 1330-1338 (2025). https://doi.org/10.1021/acsaom.5c00105

15 Aguilar-Galindo, F., Zapata-Herrera, M., Díaz-Tendero, S., Aizpurua, J. & Borisov, A. G. Effect of a Dielectric Spacer on Electronic and Electromagnetic Interactions at Play in Molecular Exciton Decay at Surfaces and in Plasmonic Gaps. *ACS Photonics* **8**, 3495-3505 (2021). https://doi.org/10.1021/acsphotonics.1c00791

16 Raja, A. *et al.* Coulomb engineering of the bandgap and excitons in two-dimensional materials. *Nature Communications* **8**, 15251 (2017). https://doi.org/10.1038/ncomms15251

17 Shin, D., Foord, J., Egdell, R. & Walsh, A. Electronic structure of CuCrO2 thin films grown on Al2O3 (001) by oxygen plasma assisted molecular beam epitaxy. *Journal of Applied Physics* **112** (2012).

18 Ok, J. M. *et al.* Interfacial stabilization for epitaxial CuCrO2 delafossites. *Scientific Reports* **10**, 11375 (2020).

19 Moreira, M., Afonso, J., Crepelliere, J., Lenoble, D. & Lunca-Popa, P. A review on the p-type transparent Cu–Cr–O delafossite materials. *Journal of Materials Science* **57**, 3114-3142 (2022).

20 Wang, X., Meng, W. & Yan, Y. Electronic band structures and excitonic properties of delafossites: A GW-BSE study. *Journal of Applied Physics* **122** (2017).

21 Elliott, R. J. Intensity of optical absorption by excitons. *Physical Review* **108**, 1384 (1957).

22 Tanguy, C. Optical dispersion by Wannier excitons. *Physical review letters* **75**, 4090 (1995).

23 Dietz, R., Hopfield, J. & Thomas, D. Excitons and the absorption edge of ZnO. *Journal of Applied Physics* **32**, 2282-2286 (1961).

24 Scanlon, D. O., Godinho, K. G., Morgan, B. J. & Watson, G. W. Understanding conductivity anomalies in CuI-based delafossite transparent conducting oxides: Theoretical insights. *The Journal of chemical physics* **132** (2010).

25 Crepelliere, J., Moreira, M., Lunca-Popa, P., Leturcq, R. & Lenoble, D. On the charge transport models in high intrinsic defect doped transparent and conducting p-type Cu–Cr–O delafossite thin films. *Journal of Physics D: Applied Physics* **58**, 015310 (2025).

26 Ok, J. M. *et al.* Pulsed-laser epitaxy of metallic delafossite PdCrO2 films. *APL Materials* **8** (2020).

27 Stroud, D. Percolation effects and sum rules in the optical properties of composities. *Physical Review B* **19**, 1783 (1979).

28 Noh, T. W. *et al.* Percolation effects in the optical properties of Ni-MgO composites. *Physical Review B* **33**, 3793 (1986).

29 Niklasson, G. A., Granqvist, C. & Hunderi, O. Effective medium models for the optical properties of inhomogeneous materials. *Applied Optics* **20**, 26-30 (1981).

30 Chen, H. *et al.* Plasmonic percolation: plasmon-manifested dielectric-to-metal transition. *ACS nano* **6**, 7162-7171 (2012).

31 Hövel, M., Gompf, B. & Dressel, M. Dielectric properties of ultrathin metal films around the percolation threshold. *Physical Review B—Condensed Matter and Materials Physics* **81**, 035402 (2010).

32 Sykes, M. F. & Essam, J. W. Exact critical percolation probabilities for site and bond problems in two dimensions. *Journal of Mathematical Physics* **5**, 1117-1127 (1964).

33 Malarz, K. Percolation thresholds on a triangular lattice for neighborhoods containing sites up to the fifth coordination zone. *Physical Review E* **103**, 052107 (2021).

34 Bogardus, E. H. & Bebb, H. B. Bound-exciton, free-exciton, band-acceptor, donor-acceptor, and auger recombination in GaAs. *Physical Review* **176**, 993 (1968).

35 He, C. *et al.* Free and bound excitonic effects in Al0. 5Ga0. 5N/Al0. 35Ga0. 65N MQWs with different Si-doping levels in the well layers. *Scientific Reports* **5**, 13046 (2015).

36 Tongay, S. *et al.* Defects activated photoluminescence in two-dimensional semiconductors: interplay between bound, charged and free excitons. *Scientific reports* **3**, 2657 (2013).

37 Crêpellière, J. *et al.* Transparent conductive CuCrO 2 thin films deposited by pulsed injection metal organic chemical vapor deposition: Up-scalable process technology for an improved transparency/conductivity trade-off. *Journal of Materials Chemistry C* **4**, 4278-4287 (2016).

38 Zhakina, E. *et al.* Investigation of Planckian behavior in a high-conductivity oxide: PdCrO2. *Proceedings of the National Academy of Sciences* **120**, e2307334120 (2023).

39 Mackenzie, A. P. The properties of ultrapure delafossite metals. *Reports on Progress in Physics* **80**, 032501 (2017).

40 Kumagai, M. & Takagahara, T. Excitonic and nonlinear-optical properties of dielectric quantum-well structures. *Physical Review B* **40**, 12359 (1989).

41 Ono, S. & Ohno, K. Minimal model for charge transfer excitons at the dielectric interface. *Physical Review B* **93**, 121301 (2016).

42 Thoai, D. T., Zimmermann, R., Grundmann, M. & Bimberg, D. Image charges in semiconductor quantum wells: Effect on exciton binding energy. *Physical Review B* **42**, 5906 (1990).

43 Jackson, J. D. & Fox, R. F. (American Association of Physics Teachers, 1999).

44 Poienar, M. *et al.* Revisiting the properties of delafossite CuCrO2: A single crystal study. *Journal of Solid State Chemistry* **185**, 56-61 (2012).

45 Giannozzi, P. *et al.* Advanced capabilities for materials modelling with QUANTUM ESPRESSO. *J Phys-Condens Mat* **29** (2017). https://doi.org/ARTN 465901
10.1088/1361-648X/aa8f79

46 Liechtenstein, A. I., Anisimov, V. I. & Zaanen, J. Density-Functional Theory and Strong-Interactions - Orbital Ordering in Mott-Hubbard Insulators. *Physical Review B* **52**, R5467-R5470 (1995). https://doi.org/DOI 10.1103/PhysRevB.52.R5467

47 Deslippe, J. *et al.* BerkeleyGW: A massively parallel computer package for the calculation of the quasiparticle and optical properties of materials and nanostructures. *Comput Phys Commun* **183**, 1269-1289 (2012). https://doi.org/10.1016/j.cpc.2011.12.006

48 Hybertsen, M. S. & Louie, S. G. Electron Correlation in Semiconductors and Insulators - Band-Gaps and Quasi-Particle Energies. *Physical Review B* **34**, 5390-5413 (1986). https://doi.org/DOI 10.1103/PhysRevB.34.5390

49 Deslippe, J., Samsonidze, G., Jain, M., Cohen, M. L. & Louie, S. G. Coulomb-hole summations and energies for
calculations with limited number of empty orbitals: A modified static remainder approach.
*Physical Review B* **87** (2013). https://doi.org/ARTN 165124

10.1103/PhysRevB.87.165124

50 Rohlfing, M. & Louie, S. G. Electron-hole excitations and optical spectra from first principles. *Physical Review B* **62**, 4927-4944 (2000). https://doi.org/DOI 10.1103/PhysRevB.62.4927

## Funding

This work was primarily sponsored by the U.S. Department of Energy, Office of Science, Basic Energy Sciences, Materials Sciences and Engineering Division. Part of optical measurement and data analysis was supported by National R&D Program through the National Research Foundation of Korea (NRF) funded by Ministry of Science and ICT (RS-2024-00348920). The initial optical characterization and data analysis at the University of Kentucky was supported by National Science Foundation Grant No. DMR-2426874. The theory work was supported by National Science Foundation (NSF) DMR-2124934. The simulation used Anvil at Purdue University through allocation DMR100005 from the Advanced Cyberinfrastructure Coordination Ecosystem: Services & Support (ACCESS) program, which is supported by NSF grants No. 2138259, 2138286, 2138307, 2137603, and 2138296.

## Author contributions

J. S., U. C., S. L., and H. N. L. conceived the idea and designed the experiments. J. S., S. L., and H. N. L. performed growth of film, characterization, and data analysis. U. C., B. K., C. S., and A. S. performed ellipsometry measurements and related analysis. D. L. and L. Y. performed theoretical calculations. J. S., U.C., C.S., and H. N. L. wrote the manuscript with contributions from all authors.

## Additional information

**Supplementary information** is available in the online version of the paper. Correspondence and requests for materials should be addressed to H. N. L.

## Competing financial interests

The authors declare no competing financial interests.

## Data availability

The data supporting the findings of this study are available from the corresponding author upon request.

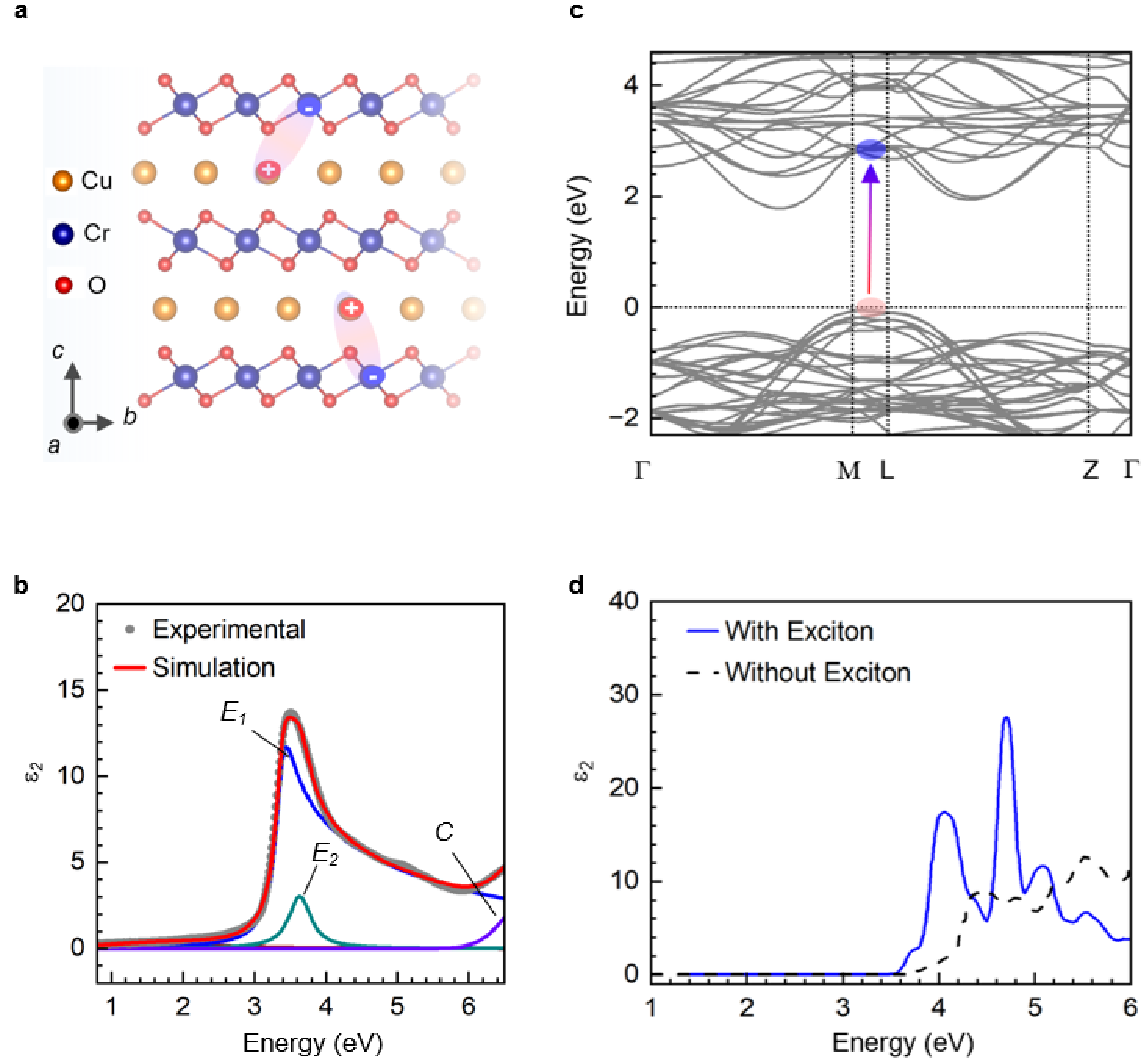

**Figure 1. Interlayer excitons in delafossite $CuCrO_2$.** **a.** Schematic of the $CuCrO_2$ crystal structure and interlayer exciton formation. The red and blue spheres represent the hole and electron, respectively. **b.** Imaginary part of the dielectric function ($\varepsilon_2$) of $CuCrO_2$ measured at room temperature (grey dotted line) and fitted using the Tanguy model (red line). The fit resolves two excitonic transitions, $E_1$ and $E_2$, associated with interlayer Cu–Cr 3*d* excitations, along with a higher-energy charge-transfer transition (*C*) from O 2*p* to Cr 3*d* states. **c.** Calculated electronic band structure of $CuCrO_2$. **d.** Comparison of the calculated imaginary part of $\varepsilon_2$ obtained from GW–BSE (blue line, including electron–hole interactions) and DFT (black dashed line, excluding electron–hole interactions).

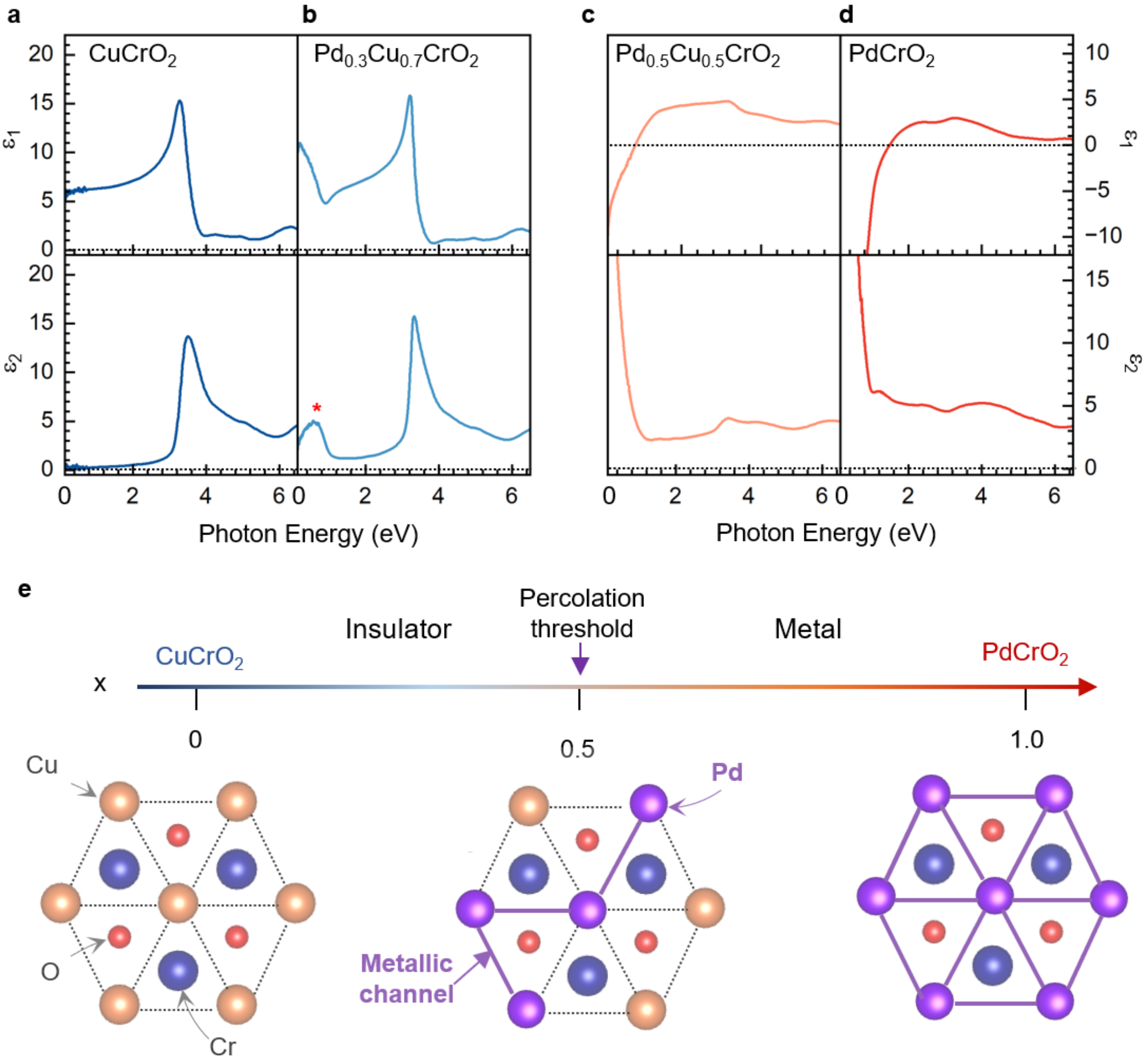

**Figure 2. Percolation-driven metal–insulator transition in $Pd_xCu_{1-x}CrO_2$. a-d,** Real ($\varepsilon_1$) and imaginary ($\varepsilon_2$) parts of the dielectric function measured at room temperature for **a.** $x = 0.0$, **b.** $x = 0.3$, **c.** $x = 0.5$, and **d.** $x = 1.0$. A broad low-energy feature (red asterisk) emerges at the intermediate composition near $x = 0.3$, consistent with localized plasmonic response. **e.** Schematic illustration of the percolation-driven metal-insulator transition. At $x \approx 0.5$, corresponding to the site-percolation threshold of a triangular lattice, conductive pathways form. For $x = 1.0$, the system becomes the fully metallic $PdCrO_2$.

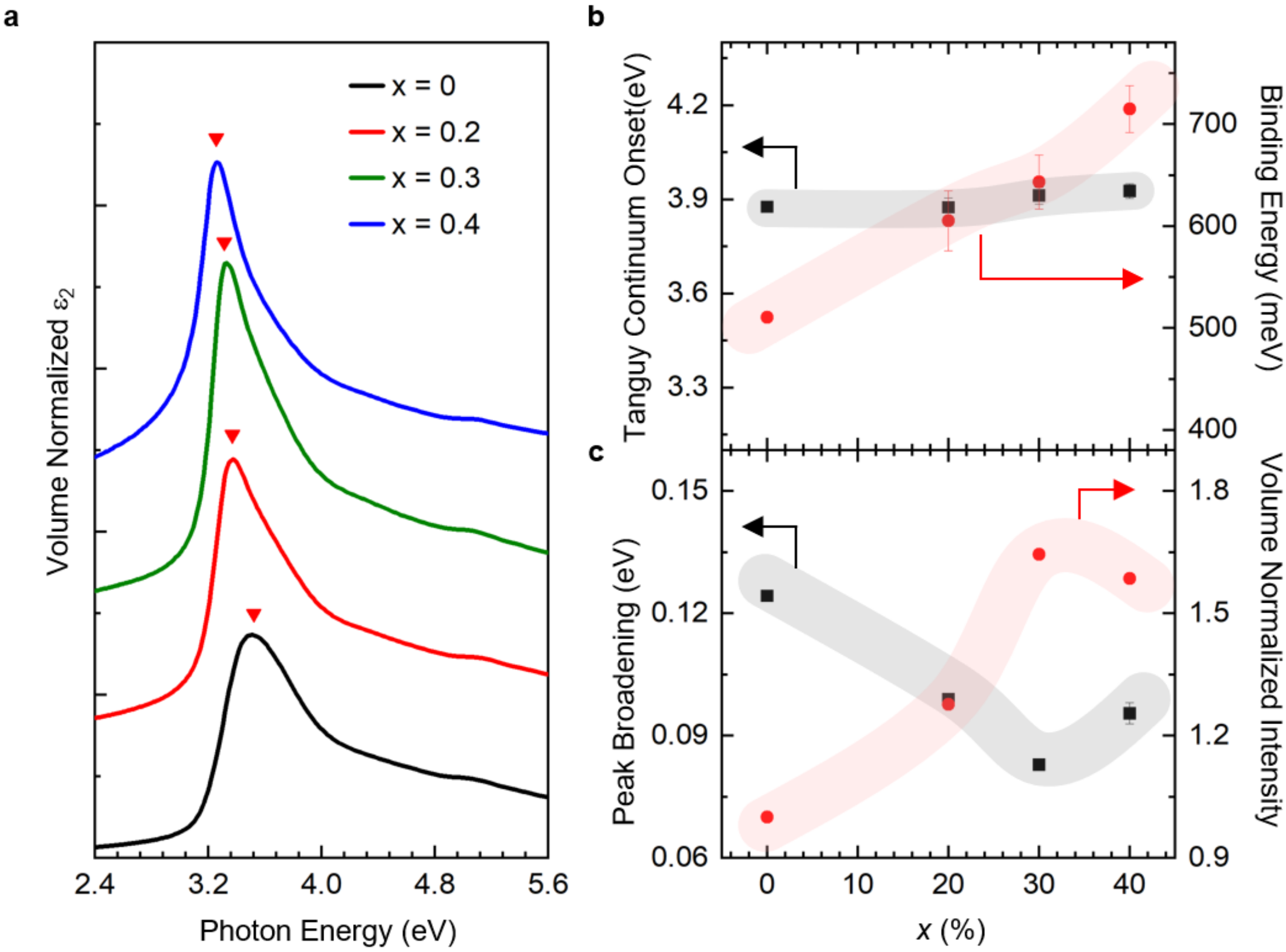

**Figure 3. Exciton stabilization in pre-percolation regime. a.** Volume-normalized imaginary part of the dielectric function ($\varepsilon_2$) for $x = 0.0$, 0.2, 0.3, and 0.4. Red inverted triangles mark the peak positions of the excitonic resonance, showing a systematic redshift with increasing Pd concentration. **b.** Evolution of the optical band gap (black squares) and exciton binding energy ($E_b$, red circles) extracted by fitting the Tanguy model as a function of Pd concentration ($x$). The binding energy increases monotonically in the pre-percolation regime. **c.** Excitonic linewidth (black squares) obtained by fitting the Tanguy model and the volume-normalized peak intensity (red circles) as a function of Pd concentration ($x$). The linewidth decreases, while the intensity increases, indicating enhanced excitonic stability.

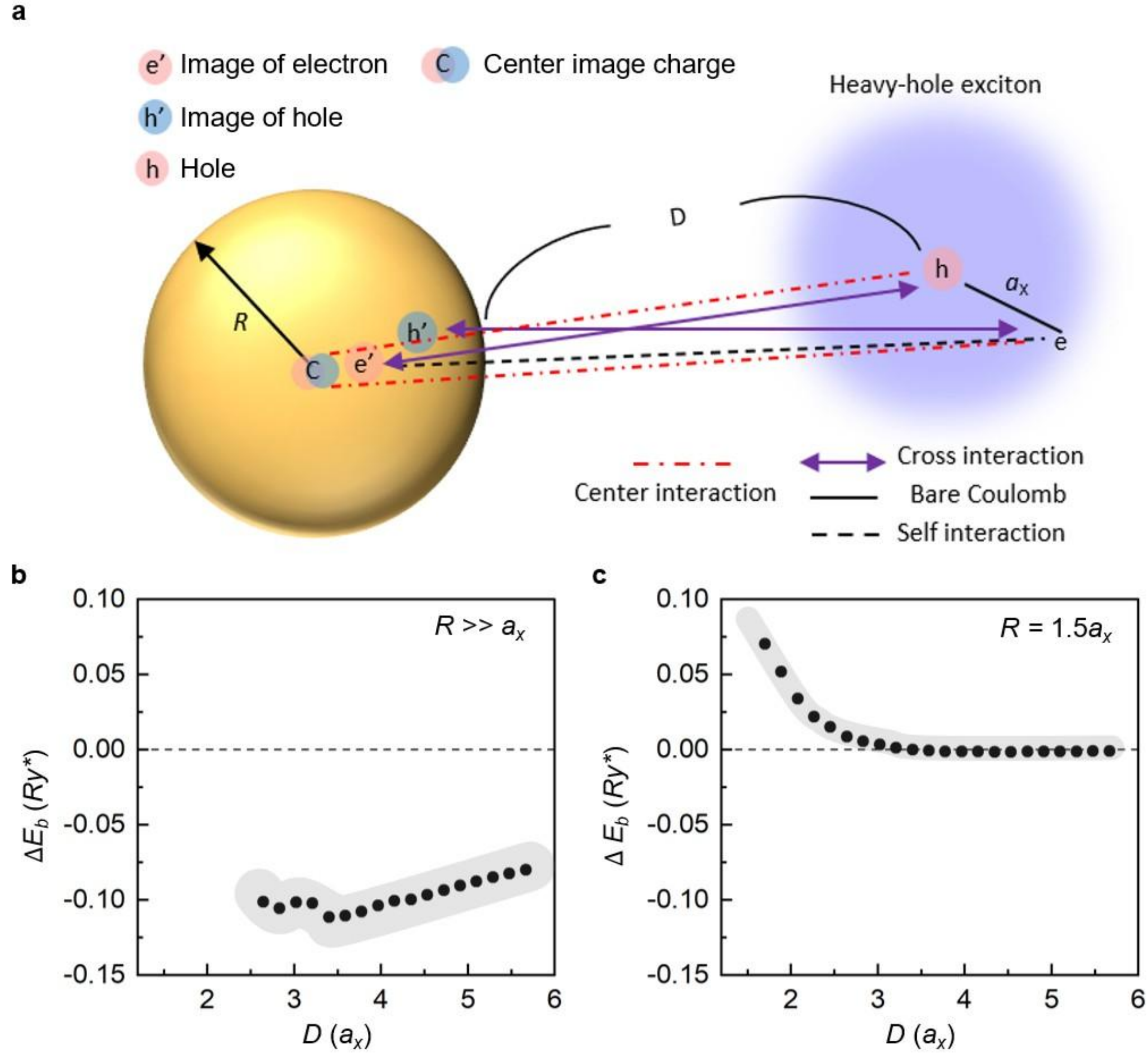

**Figure 4. Image-charge mechanism for exciton stabilization. a.** Schematic of a heavy-hole exciton interacting with an isolated, neutral metallic sphere of radius $R$. The heavy hole ($h$) is positioned at a surface-to-hole distance $D$. To satisfy the equipotential and neutrality boundary conditions, the metal induces near-surface image charges ($e'$, $h'$) and neutralizing center image charges (C). This isolated metallic sphere generates additional self-, cross-, and center-interactions that perturb the bare Coulomb potential. The exciton Bohr radius is denoted by $a_X$. **b.** Calculated change in exciton binding energy ($\Delta E_b$) as a function of separation distance ($D$) in the macroscopic metal limit (R >> $a_X$), where conventional screening reduces the binding energy. **c.** Simulated $\Delta E_b$ for a single metallic sphere ($R$ = 1.5$a_X$), showing enhancement of exciton binding when the separation distance is comparable to the exciton size.